\documentclass{article}
\usepackage{spconf,amsmath,graphicx}
\usepackage[utf8]{inputenc}
\usepackage{svg}
\usepackage{subfig}
\usepackage{verbatim}
\usepackage{url}

\title{ANALYZING THE IMPACT OF SPEAKER LOCALIZATION ERRORS ON SPEECH SEPARATION FOR AUTOMATIC SPEECH RECOGNITION}
\name{Sunit Sivasankaran, Emmanuel Vincent, Dominique Fohr \thanks{This work was made with the support of the French National Research Agency, in the framework of the project VOCADOM “Robust voice command adapted to the user and to the context for AAL” (ANR-16-CE33-0006). Experiments presented in this paper were carried out using the Grid’5000 testbed, supported by a scientific interest group hosted by Inria and including CNRS, RENATER and several Universities as well as other organizations (see https://www.grid5000.fr).}}
\address{Université de Lorraine, CNRS, Inria, LORIA, F-54000 Nancy, France}

\begin{document}
%
\maketitle
\begin{abstract}
We investigate the effect of speaker localization on the performance of speech recognition systems in a multispeaker, multichannel environment. Given the speaker location information, speech separation is performed in three stages. In the first stage, a simple delay-and-sum (DS) beamformer is used to enhance the signal impinging from the speaker location which is then used to estimate a  time-frequency mask corresponding to the localized speaker using a neural network.
This mask is used to compute the second order statistics and to derive an adaptive beamformer in the third stage. We generated a multichannel, multispeaker, reverberated, noisy dataset inspired from the well studied WSJ0-2mix and study the performance of the proposed pipeline in terms of the word error rate (WER). An average WER of $29.4$\% was achieved using the ground truth localization information and $42.4$\% using the localization information estimated via GCC-PHAT. The signal-to-interference ratio (SIR) between the speakers has a higher impact on the ASR performance, to the extent of reducing the WER by $59$\% relative for a SIR increase of $15$ dB. By contrast, increasing the spatial distance to $50^\circ$ or more improves the WER by $23$\% relative only. 
\end{abstract}

\section{Introduction}
\label{sec:intro}

Speech captured by a distant microphone is corrupted by reverberation and noise. In a typical home scenario, it is often further distorted by interfering speakers. This problem, referred to as the speech separation problem or the cocktail party problem, has been studied for more than 20 years \cite{brown_computational_1994,wang_computational_2006-1}. With the advent of neural networks, it has regained the attention of the community. 


In presence of multiple speakers different time-frequency bins are dominated by different speakers and the goal is to estimate a time-frequency mask for each speaker. The problem has been addressed in both single-channel and multichannel contexts. Single-channel approaches include clustering-based methods such as deep clustering \cite{hershey_deep_2016} and deep attractor networks \cite{chen_deep_2017}  where a neural network is trained to cluster together the time-frequency bins dominated by the same speaker. In another approach, the speakers are estimated iteratively \cite{kinoshita_listening_2018} using neural networks with permutation-invariant training criteria \cite{kolbaek_multitalker_2017-1}.

In multichannel scenarios, the usual approach is to estimate the second-order statistics (covariance matrices) of all speech and noise sources and to derive a beamformer to separate the speakers \cite{brandstein_microphone_2001,gannot_consolidated_2017-2}. The separation quality will therefore depend on the covariance matrix estimates. Different methods have been proposed to estimate the target speech and noise covariance matrices. In \cite{wang_combining_2019}, the phase differences between the microphones encoding speaker location information are exploited as input features to train a deep clustering based neural network for speech separation. Explicit speaker location estimates have also been employed. In \cite{perotin_multichannel_2018-2} and \cite{chen_multi-channel_2018}, the  speaker is first localized and the microphone signal is beamformed towards the speaker. The beamformed signal is used to estimate a mask corresponding to the localized speaker which is then used to estimate the covariance matrices. A similar approach is proposed in \cite{taseska_doa-informed_2017} where the so-called speech presence probability (SPP) is estimated using speaker location information with a minimum Bayes risk detector. The speech and noise statistics are then derived from the SPP.

Speech separation algorithms are often evaluated using speech enhancement metrics such as the signal-to-distortion ratio (SDR) and the perceptual estimation of speech quality (PESQ) metric \cite{wang_combining_2019,taseska_doa-informed_2017} and, in limited cases, using automatic speech recognition (ASR) metrics \cite{chen_cracking_2017,hendrik_barfuss_impact_2016,chen_multi-channel_2018}. There are very few  studies analyzing the impact of localization errors on WER in large vocabulary speech recognition systems. We found one closely related study in \cite{hendrik_barfuss_impact_2016}, but under limited acoustic conditions  and vocabulary size.

In this paper, we provide the following contributions. We create a new multichannel, multispeaker, reverberated, noisy dataset which extends the original WSJ0-2mix single-channel, non-reverberated, noiseless dataset \cite{hershey_deep_2016} to the strong reverberation and noise conditions and the Kinect-like microphone array geometry used in CHiME-5 \cite{barker_fifth_2018}. This allows us to use the real noise captured as part of the CHiME-5 dataset, thereby making the simulated dataset quite realistic and challenging. On this dataset, we perform speech separation using either the ground truth location of the speakers or the location estimated by the established generalized cross-correlation phase transform (GCC-PHAT) algorithm  \cite{knapp_generalized_1976}, and we evaluate the resulting ASR performance on the separated speech. To the best of our knowledge, this is the first evaluation of ASR performance on a multichannel, reverberated, noisy version of WSJ0-2mix by contrast with the ASR evaluation reported for the original single-channel, non-reverberated, noiseless WSJ0-2mix in \cite{menne_analysis_2019}.

The rest of the paper is organized as follows. Section \ref{sec:proposed} introduces the proposed framework for speech separation using speaker localization information. Section \ref{sec:dataset} explains the procedure used to simulate the dataset. Section \ref{sec:exp} describes the experimental procedure and the obtained results. We conclude in Section \ref{sec:conclusion}.

\section{Speech separation using localization information}
\label{sec:proposed}
\subsection{Signal model}
\label{subsec:model}
The multichannel signal $\mathbf{x}(t) = [x_1(t),  \dots, x_I(t)]^T$ captured at $I$ microphones can be expressed as     $   \mathbf{x}(t) = \sum_{j=1}^{J} \mathbf{c}_j(t)$, where $\mathbf{c}_j(t) = [c_{1j}(t), \dots, c_{Ij}(t)]^T$ is the spatial image of source $j$, i.e., the signal emitted by the source and captured at the microphones. Similar to \cite{gannot_consolidated_2017-2}, the  microphone index and the time index are denoted by $i$ and $t$, respectively, and $J$ is the total number of sources. This general formulation is valid for both point sources as well as diffuse noise. For point sources such as human speakers, the spatial image can be expressed as a linear convolution of the room impulse response (RIR) $\mathbf{a}_j(t, \tau) = [a_{1j}(t,\tau), \dots, a_{Ij}(t, \tau)]^T$ and a single-channel source signal $s_j(t)$: $\mathbf{c}_j(t) = \sum_{\tau=0}^{\infty} \mathbf{a}_j( t, \tau) s_j(t-\tau)$.  Under the narrowband approximation, $\mathbf{c}_j$ in the time-frequency domain can be written as $\mathbf{c}_j(t,f) = \mathbf{a}_j(t,f) s_j(t,f)$.

\begin{figure}[ht!]
\label{fig:pipeline}
\caption{Speech separation pipeline using rank-1 MWF as the adaptive beamformer.}
\includegraphics[width=\columnwidth]{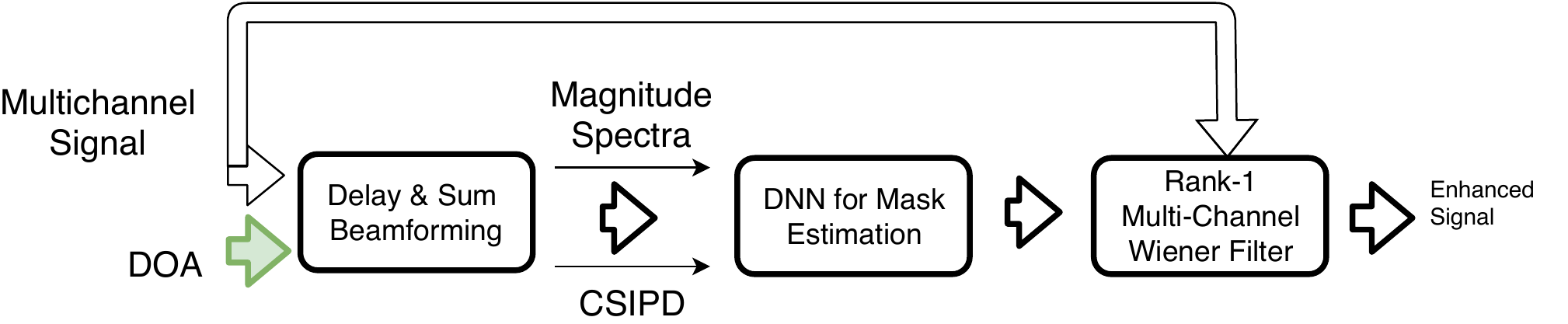}
\end{figure}

Our objective is to estimate the spatial image of each source given its (known or estimated) spatial location. An overview of our speaker location guided speech separation system is shown in Fig.~\ref{fig:pipeline}. This system comprises three steps: delay-and-sum (DS) beamforming, mask estimation, and adaptive beamforming. We detail each of these steps in the three subsections below.

\subsection{DS beamforming}
\label{subsec:doa_usage}
Given the spatial location of source $j$ in far-field, the corresponding time difference of arrival between a pair of microphones $i$ and $i^\prime$ can be obtained as:

\begin{equation}
\text{TDOA}(i,i^\prime,j)=\frac{d_{ii^\prime} \cos(\theta_{ii^\prime j})}{c}
\end{equation}
where $\theta_{ii^\prime j}$ is the direction of arrival (DOA) of the source with respect to the microphone pair $(i, i^\prime)$, $d_{ii^\prime}$ is the distance between the two microphones, and $c$ is the velocity of sound.

\begin{figure*}[t!]
\subfloat[\label{subfig:phase_diff_direct_before_ds_noise}]{\includegraphics[scale=0.3]{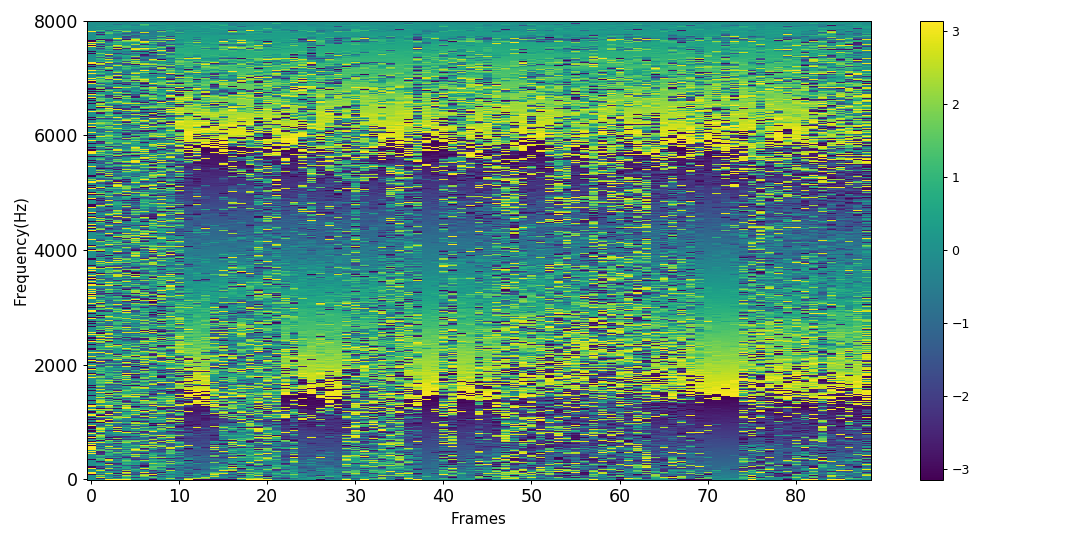}} \hfill
\subfloat[\label{subfig:phase_diff_direct_after_ds_noise}]{\includegraphics[scale=0.3]{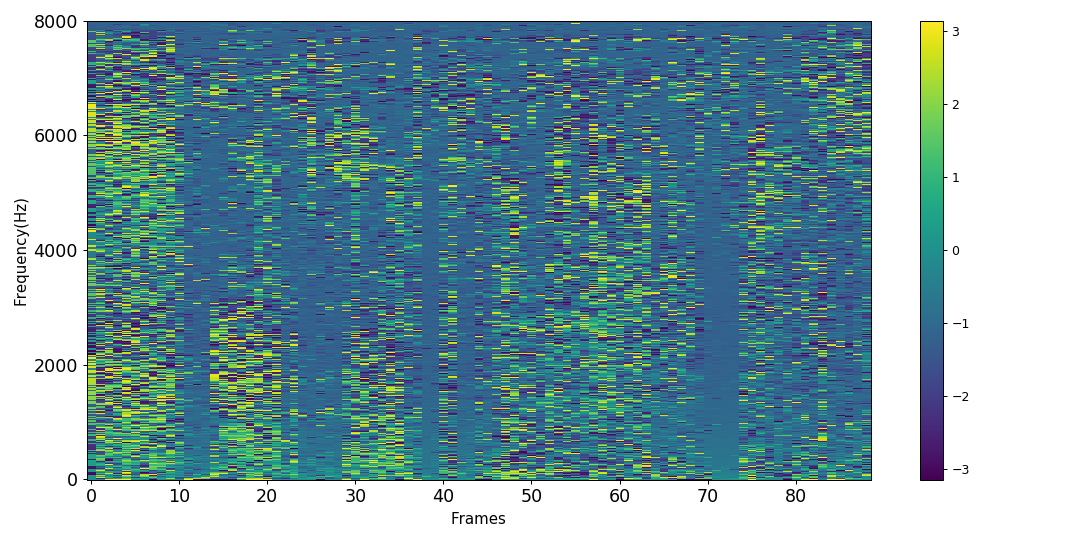}}
\caption{Phase difference in presence of noise before DS beamforming  (\ref{subfig:phase_diff_direct_before_ds_noise}) and after DS beamforming (\ref{subfig:phase_diff_direct_after_ds_noise})}
\label{fig:ds_phase}
\end{figure*}

A steering vector with respect to a reference microphone (in the following, microphone $1$) can be computed as
$\mathbf{\Tilde{d}}_j(f) = [1, e^{-2j\pi(q_{ij}-q_{1j})v_f/c}, \dots, e^{-2j\pi(q_{Ij}-q_{1j})v_f/c}]^T$
where $\nu_f = f \times f_s/F$ is the continuous frequency (in Hz), and $f_s$ is the sampling frequency. The output of a simple DS beamformer for source $j$ can then be obtained as
\begin{equation}
    \mathbf{\hat{c}}_{j,\text{DS}}(t,f) = \mathbf{\Tilde{d}}_j(f)^H \mathbf{x}(t,f).
\end{equation}
where $^H$ denotes Hermitian transposition.

The localized speaker $j$ is more prominent in $\mathbf{\hat{c}}_{j,\text{DS}}$ than in $\mathbf{x}$. We hence use $\mathbf{\hat{c}}_{j,DS}$ to compute a time-frequency mask corresponding to that speaker.

\subsection{Time-frequency mask estimation}
\label{subsec:mask_est}
The magnitude spectrum of $\mathbf{\hat{c}}_{j,DS}$ and its phase difference with respect to the reference microphone are used as inputs to a neural network that estimates the time-frequency mask corresponding to the localized speaker.

Using the phase difference between $\mathbf{\hat{c}}_{j,DS}$ and a reference microphone as a feature may not seem intuitive at first and requires further justification. Figure \ref{fig:ds_phase} shows the information captured by this phase difference. Fig.~\ref{subfig:phase_diff_direct_before_ds_noise} shows the phase difference of the direct component (without reverberation) of a source between two microphones placed at a distance of 0.226~m in the presence of noise. The phase difference is perturbed in the time-frequency bins dominated by noise.

Fig.~\ref{subfig:phase_diff_direct_after_ds_noise} shows the phase difference of the beamformed signal with respect to the signal at the reference microphone. The phase difference in the time-frequency bins dominated by speech is now zero, and a clear speech-like pattern can be observed in these bins. In the presence of reverberation, the speech patterns are less clearly visible before or after DS beamforming. Nevertheless, we argue that the phase difference contains useful information regarding the source which can be leveraged by a neural network in addition to the magnitude spectrum of the DS beamformer ouput in order to estimate a better time-frequency mask.
    
Since the phase difference is defined modulo $2\pi$ only, we use its cosine and sine as features, as used in \cite{sivasankaran_keyword-based_2018-1} for speaker localization  and in \cite{wang_combining_2019} for speech separation. We refer to these features as cosine-sine interchannel phase difference (CSIPD) features. These features are given as inputs along with the magnitude spectrum of  $\mathbf{\hat{c}}_{j,DS}$ to train a neural network to estimate a mask. We highlight the fact that the dimension of the input features to train the mask estimation network does not depend on the number of microphones. In theory, we can use the same network for any number of microphones in the array.

\subsection{Adaptive beamforming}
The mask $M_j(t,f)$ output by the neural network for a given source $j$ can be used to estimate the covariance matrix of that source as
\begin{equation}
    \mathbf{\Sigma}_j(t,f) = \alpha \mathbf{\Sigma}_j(t-1,f) + (1-\alpha) M_j(t,f) \mathbf{x}(t,f) \mathbf{x}^H(t,f)
\end{equation}
where $\alpha$ is a forgetting factor. Similarly, the noise covariance matrix $\mathbf{\Sigma}_n$, which includes the statistics corresponding to all other speakers and background noise, can be estimated as
\begin{multline}
    \mathbf{\Sigma}_\mathbf{n}(t,f) = \alpha \mathbf{\Sigma}_\mathbf{n}(t-1,f)\\ + (1-\alpha) (1-M_j(t,f)) \mathbf{x}(t,f) \mathbf{x}^H(t,f).
\end{multline}

An adaptive beamformer, i.e., a beamformer depending on the above statistics rather than the spatial location, is applied to the mixture signal $\mathbf{x}(t,f)$ to recover the sources. The output of the beamformer is $\mathbf{w}^H(t,f)\mathbf{x}(t,f)$. Different beamformers can be defined based on the chosen optimization criterion \cite{Woelfel2009,gannot_consolidated_2017-2}.  In this work we consider the generalized eigenvalue (GEV) beamformer \cite{warsitz_blind_2007}, the speech distortion weighted multichannel Wiener filter (SDW-MWF) \cite{spriet_spatially_2004-2}, and the rank-1 constrained multichannel Wiener Filter (R1-MWF) \cite{wang_rank-1_2018}.

\vspace{-3mm}
\section{Dataset}
\vspace{-2mm}
\label{sec:dataset}
The data for this work is based on a multichannel, reverberated, noisy version of the WSJ0-2mix dataset \footnote{The code to recreate the dataset can be found here: \url{https://github.com/sunits/Reverberated_WSJ_2MIX}}. The original WSJ0-2mix dataset introduced in \cite{hershey_deep_2016} was created by mixing pairs of speakers from the WSJ0 corpus, and contains 20~k, 5~k, and 3~k training, development, and test mixtures, respectively. Each mixture contains two different speakers speaking for a variable duration. In this work, the ``max'' version of the dataset is used where the length of mixed signals is the maximum of the length of individual signals.

In our experiments we emulate the recording conditions of the CHiME-5 corpus which was recorded using  Microsoft Kinect devices. For each pair of speech signals in WSJ0-2mix, we simulate room impulse responses (RIRs) using the RIR Simulator \cite{habets_rir-generator_2018} for two distinct spatial locations with a minimum DOA difference of 5$^\circ$. The room dimensions and the reverberation time (RT60) are randomly chosen in the range of $[3-9]$~m and $[0.3-1]$~s. The two speech signals are convolved with these RIRs and mixed at a random signal-to-interference ratio (SIRs) in the range of $[0-10]$~dB.

Real multichannel noise captured as part of the CHIME-5 dataset is then added with a random SNR in the range of $[0-10]$~dB. To obtain noise segments, the ground truth speech activity detection (SAD) labels from Track 3 of the DIHARD-II speaker diarization challenge \cite{ryant_second_2019} are used, as these are more reliable than the SAD labels originally provided in CHiME-5. The noise signals in the training, development, and test sets are taken from different CHiME-5 sessions.  The noise is realistic and non-stationary in nature and makes the speech separation task very challenging. A reverberated dataset based on WSJ0-2mix was reported earlier in \cite{wang_multi-channel_2018}, but it does not contain any noise.

\begin{table}[ht]
\centering
\caption{Baseline WER (\%) achieved on single-speaker or two-speaker mixtures before enhancement/separation. All results reported in this paper are with reverberated speech.}
\begin{tabular}{|c|c|c|}
\hline
\textbf{$1$ speaker} &  \textbf{$1$ speaker + noise} &  \textbf{$2$ speakers + noise} \\ \hline
12.2 &  23.6 &  58.2                   \\ \hline
\end{tabular}
\label{tab:baseline}
\end{table}

\begin{table*}[ht!]
    \caption{WER(\%) achieved on noisy two-speaker mixtures after separation using ground truth or estimated speaker location information, as a function of the DOA difference ($\Delta$ DOA) between the two speakers.}
    \resizebox{\textwidth}{!}{%
        \begin{tabular}{|l||l|l||l|l||l|l||l|l||l|l|}
            \hline
            $\Delta$ DOA & \multicolumn{2}{c|}{\textbf{\textless{}$10^\circ$}} & \multicolumn{2}{c|}{[\textbf{$10^\circ-25^\circ$}]}  & \multicolumn{2}{c|}{[\textbf{$25^\circ-50^\circ$}]}  & \multicolumn{2}{c|}{\textbf{\textgreater{}$50^\circ$}}  & \multicolumn{2}{c|}{\textbf{Average}} \\ \hline
                & \textbf{True DOA}     & \textbf{Est DOA}    & \textbf{True DOA} & \textbf{Est DOA} & \textbf{True DOA} & \textbf{Est DOA} & \textbf{True DOA}      & \textbf{Est DOA}     & \textbf{True DOA}      & \textbf{Est DOA}      \\ \hline
            \textbf{GEV}    &   38.0              &  54.5               &  31.1            &    47.0         &  32.7            & 43.1             &30.2                   &    41.5              &                 30.9  &  43.2                  \\ \hline
            \textbf{R1-MWF} &  37.4                &   53.9              &  29.8            &       46.3       &    30.7          & 42.7             & 28.8                  & 40.4                 &                 29.4  &   42.4                \\ \hline
            \textbf{SDW}    &     36.6             &  54.0              &  30.0             &   46.2          &     31.4         &      42.0        &  29.0                 &       40.9            &               29.6    & 42.4                  \\ \hline
        \end{tabular}
    }
    \label{tab:two_spk_doa}
\end{table*}

\section{Experimental settings and results}
\label{sec:exp}
\emph{DNN to estimate the mask}: Mask estimation is done in the time-frequency domain. The short time Fourier transform (STFT) of the 4-channel signal was computed using a sine window of length 100~ms and a shift of 50~ms resulting in a frequency dimension of 801. The input to the mask estimation network was of dimension $2403$: it comprises the magnitude spectrum of the DS signal, as well as the cosine and sine of the phase differences as detailed in Section \ref{subsec:mask_est}, each of which is of dimension 801. A 2-layer Bi-LSTM network was trained to estimate the mask corresponding to the reverberated component of the localized speaker. No dereverberation was performed. Adam was used as the optimizer.

\emph{ASR system}: For each separation method tested, the ASR system was trained on the enhanced training set using accurate senone alignments obtained from the underlying clean single-speaker utterances. The acoustic model (AM) was a 15-layer time-delayed neural network (TDNN) trained using the lattice-free  maximum mutual information criterion \cite{povey_purely_2016}. $40$ dimensional Mel frequency cepstral coefficients along with 100-dimensional i-vectors were used as input features.
    
\emph{Computing WER metrics for the mixture}:
In every mixture we perform ASR only for the speaker who spoke for the longest duration. This was done so that the insertion errors  corresponding to the speaker who spoke for a shorter duration do not effect the ASR performance. In a typical ASR system, this is handled by a voice activity detector / endpointing system which is not the focus of this work.

\emph{Estimating location using GCC-PHAT}:
Experiments were conducted using both ground truth DOA values  as well as using the DOA values estimated by GCC-PHAT. In the case of GCC-PHAT, peaks in the angular spectrum are assumed to correspond to the DOAs of the sources. The top two peaks are chosen and the peak which is closest to the true DOA is taken as the estimated DOA. Since GCC-PHAT works using 2 microphones, only the first and the last microphone of the array which are placed at a distance of 0.226~m are used.


{}
{}

\emph{Results}:
Table \ref{tab:baseline} shows the baseline ASR performance before separation. It can be observed that background noise and overlapping speech severely degrade performance.

Table \ref{tab:two_spk_doa} shows the ASR results obtained on noisy two-speaker mixtures after speech separation. An average WER of $29.4\%$ was obtained using the ground truth DOA, a relative improvement of $49$\% with respect to the system without source separation. This is close to the ASR performance for a single speaker with noise ($23.6$\%) as shown in Table \ref{tab:baseline}, showing that DOA information can indeed help source separation. The performance dropped to $42.4\%$ when the DOA was estimated using GCC-PHAT, indicating that erroneous DOA estimates decrease the separation quality. The DOA difference between the speakers was found to have an impact on the ASR performance. When the speakers are located at close angles with respect to the array (DOA difference $<10^\circ$), a WER of $37.4\%$ was obtained. When the speakers are well separated in space (say $>50^\circ$) the performance improved to $28.8\%$, a relative gain of $23\%$.

The SIR had an even bigger impact on the ASR performance as shown in Table \ref{tab:two_spk_sir}. SIRs below $-5$ dB resulted in a WER of $59.7\%$ which improved to $24.3\%$ when the SIR was above $10$ dB, a relative improvement of $59\%$.

Finally, note that the R1-MWF beamformer outperformed the widely used GEV beamformer in all our experiments.

{}
\begin{table}[ht!]
\centering
\caption{WER(\%) achieved on noisy two-speaker mixtures after separation using ground truth speaker location information, as a function of the SIR.}
\begin{tabular}{|l|l|l|l|l|l|}
\hline
\textbf{SIR} (dB)& \textless{} $-5$ & $[-5, 0]$ &  $[0, 5]$ & $[5 , 10]$ & $>10$ \\ \hline \hline
\textbf{GEV}       & 64.0         & 35.4  & 26.3   & 23.3    & 25.6        \\ \hline
\textbf{R1-MWF}    &  59.7      & 33.8   & 25.4   & 22.4  &  24.3          \\ \hline
\textbf{SDW}       & 60.2         & 33.5  & 25.8   & 22.7   & 24.7        \\ \hline
\end{tabular}
\label{tab:two_spk_sir}
\end{table}

\section{Conclusion and future work}
\label{sec:conclusion}
We conducted the first analysis of the impact of speaker localization accuracy on speech separation performance in challenging two-speaker, reverberated, noisy scenarios, as measured by the resulting ASR performance. To do so, we created a new dataset by reverberating WSJ0-2mix and mixing it with real CHiME-5 noise, and made the corresponding code publicly available. We found that the ASR performance depends more on the SIR of the speakers, with lower WERs for signals with higher SIR. The angular distance between the DOAs of the speakers was also found to have an impact, with better WERs for signals whose speakers exhibit a larger difference in DOAs. As future work we would like to investigate methods to make the mask estimation network robust to localization errors.

\newpage
\small
\bibliographystyle{IEEEbib}
\bibliography{ref}
\end{document}